\documentclass[12pt]{amsart}
\usepackage{amsmath,amstext,amsfonts,amsthm,amssymb}
\usepackage{eucal}
\usepackage{graphicx}
\renewcommand{\subsubsection}[1]{\addtocounter{subsubsection}{1}
{\ \\[3pt]\bf \thesubsubsection. \  #1} }

\swapnumbers

{  \theoremstyle{definition}

}
\newcommand{\Gal}{\operatorname{Gal}}

\newcommand{\lra}{\longrightarrow}

\newcommand{\iso}{\overset\sim\longrightarrow}

\newcommand{\bea}{\begin{eqnarray*}}
\newcommand{\eea}{\end{eqnarray*}}
\newcommand{\bean}{\begin{eqnarray}}
\newcommand{\eean}{\end{eqnarray}}

\newcommand{\tchi}{\tilde\chi}
\newcommand{\tf}{\tilde f}

\newcommand{\fp}{\mathfrak p}

\newcommand{\CW}{\mathcal{W}}

\newcommand{\BC}{\mathbb{C}}

\newcommand{\BF}{\mathbb{F}}

\newcommand{\BQ}{\mathbb{Q}}
\newcommand{\BR}{\mathbb{R}}
\newcommand{\BZ}{\mathbb{Z}}

\newcommand{\bbn}{\text{{\bf n}}}

\begin{document}


\centerline{PRODUITS GAMMA ET VECTEURS PROPRES} 

\bigskip\bigskip

\centerline{DES MATRICES DE CARTAN}


\vspace{1.5cm}

\centerline{V\'eronique Cohen-Aptel et Vadim Schechtman}

\vspace{1.5cm}

\centerline{\bf \S 1. Introduction}

\bigskip\bigskip

Le but de cet article est d'exprimer certains vecteurs propres de matrices de Cartan en termes 
de produits de valeurs de la fonction $\Gamma$. 

Pour \'enoncer le r\'esultat, fixons les notations standards sur les syst\`emes de racines, 
cf. [B].
 
Soient $V$ un espace vectoriel r\'eel de dimension $r\geq 1$, $R\subset V$ un syst\`eme de racines fini irr\'eductible. Munissons $V$ d'un produit scalaire $(.|.)$ $W$-invariant, $W$ \'etant 
le groupe de Weyl de $R$ ; \`a l'aide de ce produit $V$ sera identifi\'e  \`a son dual $V^*$ ; en particulier 
on peut consid\'erer les racines duales comme les \'el\'ements de $V$: $\alpha^\vee = 2\alpha/(\alpha,\alpha),\ \alpha\in R$. 
 
Choisissons une base $\{\alpha_i,\ 1\leq i \leq r\}$ de racines simples de $R$. 

Soit $A =  ((\alpha_i|\alpha_j^\vee))$ la matrice de Cartan de $R$ ; suivant l'usage, on dira que $R$ est simplement lac\'e si $A$ est sym\'etrique.  

Soit $\rho$ la demi-somme des racines positives; on pose : 
$$\rho^\vee = \frac{1}{2}\sum_{\alpha > 0}\ \alpha^\vee.$$
Soit $h$ le nombre de Coxeter de $R$. 

On d\'efinit des nombres r\'eels strictement positives :
$$
\Gamma(R,\alpha_i) = \prod_{\alpha > 0}\ \Gamma((\alpha|\rho^\vee)/h)^{-(\alpha^\vee|\alpha_i)},
$$
et un vecteur
$$
\Gamma(R) = (\Gamma(R,\alpha_1),\ldots, \Gamma(R,\alpha_r)) \in \BR^r.
$$

D' autre part, la matrice 
$A' = I - A/2$ est ind\'ecomposable avec des coefficients positifs (c'est la matrice d'incidence du graphe 
de Dynkin de $R$). D'apr\`es le th\'eor\`eme de Perron - Frobenius (cf. [G], Ch. XIII, \S 2), $A'$ admet un 
unique, \`a proportionalit\'e pr\`es, vecteur propre $v_{PF}$ de coordonn\'ees strictement positives et de valeur propre r\'eelle $\lambda_{max}(A') > 0$. 
Celle-ci est strictement sup\'erieure aux valeurs absolues de toutes autres valeurs propres de $A'$.  

En effet toutes les valeurs propres de $A$ sont re\'elles et strictement positives, et $v_{PF}$ est un vecteur 
propre de $A$ de valeur propre minimale :
$$
\lambda_{min}(A) = 4\sin^2(\pi/2h),
$$
o\`u $h$ est le nombre de Coxeter de $R$. On a 
$$
\lambda_{max}(A') = (2 - \lambda_{min}(A))/2 = \cos(\pi/2h).
$$
On appelle $v_{FP}$ le {\it vecteur de Perron-Frobenius} de $A$. 

L'\'enonc\'e suivant est le r\'esultat principal de cette note. 

{\bf 1.1.} {\it Th\'eor\`eme.} Pour chaque syst\`eme de racines $R$ fini irr\'eductible de rang $r$, 
tous les nombres $\pi\Gamma(R,\alpha_i),\ 1\leq i \leq r$, sont alg\'ebriques.  

Le vecteur  
$\Gamma(R)$ est le vecteur de Perron - Frobenius de la matrice de Cartan $A$ de $R$. 

\bigskip

Pour la preuve (qui est un calcul direct), voir \S 2 et Appendice (pour le cas $E_8$) ci-dessous. 

Soit 
$$
\theta = \sum_{i=1}^r\ n_i\alpha_i
$$
la plus longue racine ; on pose $n_0 = 1$, $\alpha_0 = - \theta$, donc $h = \sum_{i=0}^r\ n_i$. 

On peut aussi d\'efinir les coordonn\'ees de $v_{PF}$ comme  des {\it valeurs propres} 
d'une certaine matrice $M$ li\'ee \`a $R$, 
cf. [Fr], [FLO]. Ces nombres sont les masses des particules dans des mod\`eles massives int\'egrables, 
les th\'eories de Toda affines, cf. [Z], [FZ]. 

Maintenant on va \'enoncer une assertion semblable au th\'eor\`eme ci-dessus pour des matrices de Cartan affines. 
Soit $\hat A$ la matrice de Cartan $(r+1)\times (r+1)$ du syst\`eme de racines affine $R^{(1)}$, cf. [K]. Alors $\hat A$ admet  \`a proportionalit\'e pr\`es, un seul vecteur propre de valeur propre $0$, et 
$$
\delta := (n_0,\ldots, n_r) \in \BR^{r+1} 
$$
est l'unique tel vecteur dont les coordonn\'ees sont strictement positives et
enti\`eres. Les nombres $n_i, 0\leq i\leq r$ co\"\i ncident avec les marques de Kac du graphe de Dynkin complet\'e $R^{(1)}$ (cf. la planche Aff 1, cf. [K], 4.8, 4.9). 

Supposons d'abord que $R$ soit simplement lac\'e, i.e. du type $A_n, D_n$ ou $E_n, n = 6, 7, 8$. 

Posons $\gamma(x) = \Gamma(x)/\Gamma(1-x)$, 
$$
\gamma(R,\alpha_i) = \prod_{\alpha > 0}\ \gamma((\alpha|\rho)/h)^{-(\alpha_i|\alpha)},\ 0\leq i\leq r,
$$
$$
\gamma(R) = (\gamma(R,\alpha_0),\ldots,\gamma(R,\alpha_r))\in \BR^{r+1}.
$$
L'\'enonc\'e suivant est \'equivalent \`a une formule de V.Fateev, cf. [F1], (66). Cette formule 
remarquable a \'et\'e le point de d\'epart de cette note.

{\bf 1.2.} {\it Th\'eor\`eme.} 
$$
\gamma(R) = k(R)^{-1/h}\delta
$$
o\`u
$$
k(R) = \prod_{i=1}^r\ n_i^{n_i}.
$$ 

Pour le cas g\'en\'eral, introduisons les nombres
$$
n_i^\vee = n_i(\alpha_i|\alpha_i)/2
$$
Au coefficient de proportionalit\'e commun pr\`es, ils co\"\i ncident avec   
les marques du graphe de Dynkin dual, obtenu en renversant les arr\^ets du graphe initial. 

Soit 
$$
h^\vee = h^\vee(R) = \sum_{i=0}^r\ n^\vee_r
$$

Posons :
$$
\delta^\vee := (n_0^\vee,\ldots,n_r^\vee),
$$
c'est un vecteur propre, de  valeur propre $0$, de la matrice de Cartan g\'en\'eralis\'ee $\hat A^\vee$ 
duale de $\hat A$. 

On d\'efinit : 
$$
\gamma(R,\alpha_i) = \prod_{\alpha>0}\ \gamma((\alpha|\rho^\vee)/h)^{-(\alpha_i|\alpha^\vee)},\ 
$$
$$
\gamma(R) = (\gamma(R,\alpha_0),\ldots,\gamma(R,\alpha_r)).
$$
L'\'enonc\'e suivant est \'equivalent \`a une formule de Fateev avec collaborateurs, [ABFKR], (46). 

{\bf 1.3.} {\it Th\'eor\`eme.} 
$$
\gamma(R) = \biggl(\prod_{i=0}^r\ n_i^{\vee n_i}\biggr)^{-1/h}\delta^\vee.
$$

La preuve de 1.2 et 1.3 se trouve dans \S 3. 

Dans le dernier \S 4 on ajoute quelques remarques arithm\'etiques. On montre comment nos calculs, joints aux   
aux r\'esultats de Deligne sur les sommes de Jacobi (g\'en\'eralis\'ant les th\'eor\`emes classiques d'Andr\'e Weil) permettent \`a d\'efinir, \`a partir de $R$ muni d'une base $B$ de racines simples, des caract\`eres galoisiens
$$
\Gal(\bar\BQ/\BQ(\mu_h)) \lra \BQ(\mu_h)^* 
$$
Il s'en suit que nos produits $\Gamma$ donnent lieu au vecteurs propres en sens Galoisien: ils  
sont des vecteurs propres communs des Frobeniuses g\'eom\'etriques du corps 
$\BQ(\mu_h)$, avec les sommes de Jacobi correspondants comme les valeurs propres.   

\`A la fin on note une analogie avec un th\'eor\`eme d'Aomoto 
sur les int\'egrales de Selberg.   

\bigskip

Le deuxi\`eme auteur est reconaissant \`a  Max-Planck-Institut f\"ur Mathematik \`a Bonn, o\`u la partie 
de ce travail a \'et\'e faite en juillet 2010. Il remercie aussi M.Gorelik pour une discussion pr\'ecieuse 
\`a MPI en juillet 2011.

\bigskip\bigskip



\centerline{\bf \S 2. Preuves: cas fini}

\bigskip\bigskip

Les vecteurs de Perron-Frobenius de toutes les matrices de Cartan finis sont connus explicitement. 
Voici leurs liste. Pour les syst\`emes simplement lac\'es, voir [GHJ], [KM], [P], [F2], [BCDS]. 
 
Pour les cas non-simplement lac\'es, voire [BCDS]; dans cet article un proc\'ed\'e de pliure ("folding") 
est d\'ecrit, qui permet d'obtenir le vecteur PF (dit aussi "de masses") des syst\`emes non-simplement 
lac\'es \`a partir des systemes simplement lac\'es convenables. Ce pliure signifie que pour 
$R = B_n, C_n, F_4$ ou $G_2$, le graphe de Dynkin dual de $R^{(1)}$ appartient \`a Aff$^{(i)}$ avec 
$i = 2, 3$. Par exemple, le dual de $F_4^{(1)}$ est $E_6^{(2)}$ et le dual de $G_2^{(1)}$ est 
$D_4^{(3)}$.    

On utilise la notation suivante pour un vecteur de Perron-Frobenius d'un syst\`eme de racines $R$:  
$$
m(R) = (m_1, \ldots, m_r).
$$
On utilise la num\'erotation des sommets du graphe de Dynkin comme dans le livre de Bourbaki [B], 
Planches \`a la fin du livre. 

$A_n$: 
$$
m_a = \sin(\pi a/(n+1)),\ 1\leq a \leq n.
$$

$B_n$: 
$$
m_a = 2\sin(\pi a/2n),\ 1\leq a \leq n - 1,\ m_n = 1,
$$
cf. $D_{n+1}$. 

$C_n$: 
$$
m_a = \sin(\pi a/2n),\ 1\leq a \leq n,
$$
cf. $A_{2n-1}$. 

$D_n$: 
$$
m_a = 2\sin(\pi a/(2n-2)),\ 1\leq a \leq n - 2,\ m_{n-1} = m_n = 1.
$$

$E_6$: 
$$
m_1 = m_6 = 1,\ m_3 = m_5 = \frac{\sqrt 3 + 1}{\sqrt 2},\ 
m_4 = \sqrt 3 + 1,\ m_2 = \sqrt 2.
$$

$E_7$: 
$$
m_7 = 1,\ m_6 = 2\cos(\pi/18),\ m_1 = 2\cos(5\pi/18),\ 
m_4 = 4\cos(\pi/18)\cos(\pi/9), 
$$
$$
m_3 = 4\cos(\pi/18)\cos(5\pi/18),\ m_5 = 2\cos(\pi/9)\cos(2\pi/9),\ m_2 = 2\cos(\pi/9).
$$

$E_8$:
$$
m(E_8) = (2\cos(\pi/5), 4\cos(\pi/5)\cos(7\pi/30), 4\cos(\pi/5)\cos(\pi/30), 
$$
$$ 
8\cos^2(\pi/5)\cos(2\pi/15), 8\cos^2(\pi/5)\cos(7\pi/30), 4\cos(\pi/5)\cos(2\pi/15), 
2\cos(\pi/30), 1). 
$$

$F_4$:
$$
m_1 = \sqrt 2,\ m_2 = \sqrt 3 + 1 ,\ m_3 = \frac{\sqrt 3 + 1}{\sqrt 2},\ m_4 = 1, 
$$
cf. $E_6$. 

$G_2$:
$$
m_1 = 1,\ m_2 = \sqrt 3,
$$
cf. $D_4$. 

\bigskip

Maintenant on va calculer les vecteurs $\Gamma(R)$. Rappelons les identit\'es classiques satisfaites par  
la fonction $\Gamma(x)$: 
$$
\Gamma(x)\Gamma(1 - x) = \frac{x}{\sin(\pi x)}
\eqno{(C)}
$$
(formule des compl\'ements);
$$
\prod_{i=0}^{n-1} \Gamma(x + i/n) = (2\pi)^{(n-1)/2}n^{-nx + 1/2}\Gamma(nx)
\eqno{(M)}
$$
On aura besoin de trois cas $n = 2, 3, 5$ de cette formule; voici les cas $n = 2$ et $3$ explicitement:  
$$
\Gamma(x)\Gamma(x + 1/2) = \pi^{1/2}2^{-2x + 1}\Gamma(2x)
\eqno{(D)}
$$
(formule de duplication de Legendre) et    
$$
\Gamma(x)\Gamma(x + 1/3)\Gamma(x + 2/3) = 2\pi3^{-3x + 1/2}\Gamma(3x)
\eqno{(T)}
$$
On a
$$
\gamma(x) = \frac{\Gamma(x)}{\Gamma(1-x)} = \frac{\sin(\pi x)}{\pi}\Gamma(x)^2,
$$
donc
$$
\Gamma(x)^2 = \frac{\pi}{\sin(\pi x)}\gamma(x).
$$
On pose
$$
s(x) := \frac{\pi}{\sin(\pi x)}.
$$
Par exemple $s(1/2) = 1/\pi$. 
$$
s(R,\alpha_i) := \prod_{\alpha > 0} s((\alpha|\rho)/h)^{-(\alpha|\alpha_i)}, 
$$
d'o\`u
$$
\Gamma(R,\alpha_i)^2 = s(R,\alpha_i)\gamma(R,\alpha_i).
$$

\bigskip\bigskip

\centerline{\it Syst\`eme de racines $A_n$}

\bigskip\bigskip

Le nombre de Coxeter $h = n+1$. 

Pour $1 \leq a \leq n$ :
$$
\Gamma(A_n,\alpha_a) = \{\Gamma(a/(n+1))\Gamma((n+1-a)/(n+1))\}^{-1} = \frac{\sin(\pi a/(n+1))}{\pi}.
$$
Donc 
$$
\Gamma(A_n) 
= \pi^{-1}(\sin(\pi/(n+1)), \ldots, \sin(\pi n/(n+1)) = \pi^{-1}m(A_n).
$$

\bigskip\bigskip

\centerline{\it Syst\`eme de racines $B_n,\ n\geq 2$}

\bigskip\bigskip

$h = 2n$. Pour $1\leq a \leq n-1$ :
$$
\Gamma(B_n,\alpha_a) = 
$$
$$
\frac{\Gamma((n-a)/2n)\Gamma((2n-2a+2)/2n)}
{\Gamma(a/2n)\Gamma((2n-2a)/2r)\Gamma((n-a+1)/2n)\Gamma((2n-a+1)/2n)}
= \frac{\sin(\pi/2n)}{2^{-1/n}\pi}.
$$
(on utilise (D) avec $x = (n-a)/2n$ et $x = (n-a+1)/2n$).
$$
\Gamma(B_n,\alpha_n) = \Gamma(1/2n)^{-1}\Gamma(2/2n)\Gamma(n/2n)^{-1}\Gamma((n+1)/2n)^{-1} = 2^{-(n-1)/n}\pi^{-1}.
$$
Il s'en suit : 
$$
\Gamma(B_n) = 2^{1/n}\pi^{-1}(\sin(\pi/2n),...,\sin((n-1)\pi/2n),1/2) = 2^{1/n}\pi^{-1}m(B_n).
$$

\bigskip

\centerline{\it Syst\`eme de racines $C_n,\ n\geq 2$}

\bigskip

$h = 2n$. Pour $1\leq a \leq n$ : 
$$
\Gamma(C_n,\alpha_a) = \Gamma(a/2n)^{-1}\Gamma((2n-a)/2n)^{-1} = \frac{\sin(\pi a/2n)}{\pi}, 
$$
d'o\`u
$$
\Gamma(C_n) = \pi^{-1}(\sin(\pi/2n),\ldots,\sin(\pi n/2n) = \pi^{-1}m(C_n).
$$

\bigskip

\centerline{\it Syst\`eme de racines $D_n,\ n\geq 3$}

\bigskip

$h = 2n - 2$.
Pour $1\leq a \leq n - 2$ : 
$$
\Gamma(D_n,\alpha_a) = 
$$
$$
\frac{\Gamma((n-a-1)/(2n-2))\Gamma((2n-2a)/(2n-2))}
{\Gamma((2n-2a-2)/(2n-2))\Gamma((2n-a-1)/(2n-2))\Gamma((n-a)/(2n-2))\Gamma(a/(2n-2))} = 
$$
$$
= 
2^{1/(n-1)}\pi^{-1}\sin(\pi a/(2n-2)), 
$$
et
$$
\Gamma(D_n,\alpha_{n-1}) = \Gamma(D_n,\alpha_n) = 
$$
$$
= \Gamma(1/(2n-2))^{-1}\Gamma(2/(2n-2))
\Gamma((n-1)/(2n-2))^{-1}\Gamma(n/(2n-2))^{-1} = 2^{-n/(n-1)}\pi^{-1}, 
$$
D'o\`u :
$$
\Gamma(D_n) = 2^{1/(n-1)}\pi^{-1}m(D_n).
$$

\bigskip\bigskip

\centerline{\it Syst\`eme de racines $E_6$}

\bigskip\bigskip

$h=12$. Le graphe de Dynkin admet un automorphisme $\sigma$ d'ordre $2$, 

$\sigma(\alpha_1) = \alpha_6$, $\sigma(\alpha_3) = \alpha_5$, $\sigma(\alpha_i) = \alpha_i$ pour $i=2, 4$. 

Il s'en suit que $f(E_6,\alpha_i) = f(E_6,\sigma(\alpha_i))$ pour $f = \Gamma, \gamma$ ou $s$.

On va utiliser les formules \'el\'ementaires suivantes : 
$$
\sin(\pi/12) = \sin(\pi/3 - \pi/4) = \frac{\sqrt 3 - 1}{2}\cdot \frac{\sqrt 2}{2}.
$$
$$
\sin(5\pi/12) = \sin(7\pi/12) =\sin(\pi/3 + \pi/4) = \frac{\sqrt 3 + 1}{2}\cdot \frac{\sqrt 2}{2}.
$$
$$
s(1/4) = \sqrt 2\pi,\ s(1/3) = s(2/3) = 2\pi/\sqrt 3.
$$
$$
s(1/12)/s(5/12) = (\sqrt 3 + 1)/(\sqrt 3 - 1) = (\sqrt 3 + 1)^2/2.
$$

En plus, on utilisera le calcul du vecteur $\gamma(E_6)$, cf. ci-dessous. 

On a:
$$
\gamma(E_6,\alpha_1) = \gamma(E_6,\alpha_6) = \frac{\gamma(3/12)}{\gamma(1/12)\gamma(8/12)} = 
2^{-1/2}3^{-1/4},
$$ 
et
$$
s(E_6,\alpha_1) = s(E_6,\alpha_6) = \frac{s(1/4)}{s(1/2)s(1/12)s(2/3)} = \frac{(\sqrt 3 - 1)\sqrt 3}{2^2\pi^2}.
$$
Il s'en suit:  
$$
\Gamma(E_6,\alpha_1) = \Gamma(E_6,\alpha_6) = \frac{\Gamma(3/12)}{\Gamma(6/12)\Gamma(1/12)\Gamma(8/12)} = 
2^{-5/4}\pi^{-1} \cdot 3^{1/8}(\sqrt 3 - 1)^{1/4}.
$$

\bigskip

$$
\Gamma(E_6,\alpha_3) = \Gamma(E_6,\alpha_5) = \frac{\Gamma(6/12)}{\Gamma(4/12)\Gamma(5/12)\Gamma(9/12)}.
$$
$$
s(E_6,\alpha_3) = s(E_6,\alpha_5) = \frac{s(1/2)}{s(1/3)s(5/12)s(3/4)} = \pi^{-2}\cdot 
\frac{\sqrt 3(\sqrt 3 + 1)}{2^3}.
$$
$$
\gamma(E_6,\alpha_3) = \gamma(E_6,\alpha_5) = \{\gamma(4/12)\gamma(5/12)\gamma(9/12)\}^{-1} = 
2^{1/2}3^{-1/4}.
$$
$$
s(E_6,\alpha_1)/s(E_6,\alpha_3) = s(1/2)^{-2} s(1/4)^{2} s(5/12)/s(1/12) = 2^{2}(\sqrt 3 + 1)^{-2}.
$$
$$
\Gamma(E_6,\alpha_1)/\Gamma(E_6,\alpha_3) = \frac{\sqrt 2}{\sqrt 3 + 1}.
$$

\bigskip

$$
\Gamma(E_6,\alpha_2) = \frac{1}{\Gamma(6/12)}\cdot\frac{\Gamma(2/12)\Gamma(10/12)}{\Gamma(1/12)\Gamma(11/12)}\cdot 
\frac{\Gamma(3/12)}{\Gamma(4/12)\Gamma(5/12)}.
$$
$$
\gamma(E_6,\alpha_2) = \frac{\gamma(3/12)}{\gamma(4/12)\gamma(5/12)} = \gamma(E_6,\alpha_3) = 
2^{1/2}3^{-1/4}.
$$
$$
\Gamma(E_6,\alpha_2)/\Gamma(E_6,\alpha_3) = \frac{\Gamma(2/12)\Gamma(10/12)\Gamma(3/12)\Gamma(9/12)}{\Gamma(6/12)^2\Gamma(1/12)\Gamma(11/12)}
 = \frac{2}{\sqrt 3 + 1}.
$$

$$
\Gamma(E_6,\alpha_4) = \Gamma(1/12)\Gamma(2/12)^{-1}\Gamma(10/12)^{-1}\Gamma(3/12)^{-2}\Gamma(9/12)
\Gamma(4/12)\times
$$
$$
\Gamma(5/12)\Gamma(7/12)^{-1}\Gamma(6/12)^{-1}.
$$
$$
s(E_6,\alpha_4) = s(1/12)s(1/6)^{-2}s(1/4)^{-1}
s(1/3)s(1/2)^{-1} = \{\sqrt 3(\sqrt 3 - 1)\}^{-1}.
$$
$$
s(E_6,\alpha_4)/s(E_6,\alpha_3) = 2^{2}/3.
$$
$$
\gamma(E_6,\alpha_4) = \gamma(1/12)\gamma(3/12)^{-3}.
\gamma(4/12) = 2^{-1/2}3^{3/4}.
$$
$$
\gamma(E_6,\alpha_4)/\gamma(E_6,\alpha_3) = 3/2.
$$
$$
\Gamma(E_6,\alpha_4)/\Gamma(E_6,\alpha_3) = 2^{1/2}.
$$

Il s'en suit : 
$$
\Gamma(E_6) 
= \Gamma(E_6,\alpha_1)\cdot (1,\sqrt 2,(\sqrt 3 + 1)/\sqrt 2,\sqrt 3 + 1,(\sqrt 3 + 1)/\sqrt 2,1) = 
$$
$$
= 2^{-5/4}3^{1/8}(\sqrt 3 - 1)^{1/4}\pi^{-1}m(E_6).
$$

\bigskip

\centerline{\it Syst\`eme de racines $E_7$}

\bigskip

$h = 18$. En utilisant (D) et (T), on arrive aux valeurs suivantes de $\Gamma(E_7,\alpha_i) : $ 

$$
\Gamma(E_7,\alpha_1) = \frac{\Gamma(3/18)\Gamma(5/18)\Gamma(16/18)}
{\Gamma(1/18)\Gamma(6/18)\Gamma(8/18)\Gamma(10/18)\Gamma(17/18)}
= \frac{\sin(\pi/9)\sin(2\pi/9)}{2^{-10/9}3^{1/6}\pi}.
$$
$$
\Gamma(E_7,\alpha_2) = \frac{\Gamma(2/18)\Gamma(3/18)\Gamma(10/18)}
{\Gamma(1/18)\Gamma(5/18)\Gamma(6/18)\Gamma(7/18)\Gamma(14/18)}
= \frac{\sin(2\pi/9)}{2^{-1/9}3^{1/6}\pi}.
$$
$$
\Gamma(E_7,\alpha_3) = \frac{\Gamma(8/18)\Gamma(15/18)}{\Gamma(5/18)\Gamma(9/18)\Gamma(11/18)\Gamma(16/18)}
= 2^{-8/9}3^{1/3}/\pi.
$$
$$
\Gamma(E_7,\alpha_4) = \frac{\Gamma(1/18)\Gamma(5/18)\Gamma(9/18)\Gamma(14/18)}
{\Gamma(2/18)\Gamma(3/18)\Gamma(7/18)\Gamma(8/18)\Gamma(12/18)\Gamma(15/18)} 
= 
$$
$$
= \frac{\sin(2\pi/9)\sin(4\pi/9)}{2^{-10/9}3^{1/6}\pi}.
$$
$$
\Gamma(E_7,\alpha_5) = \frac{\Gamma(2/18)\Gamma(6/18)\Gamma(7/18)\Gamma(12/18)}
{\Gamma(3/18)\Gamma(4/18)\Gamma(5/18)\Gamma(9/18)\Gamma(10/18)\Gamma(14/18)} = 
$$
$$
= \frac{\sin(2\pi/9)\sin(5\pi/18)}{2^{-10/9}3^{1/6}\pi}.
$$
$$
\Gamma(E_7,\alpha_6) = \frac{\Gamma(3/18)}{\Gamma(2/18)\Gamma(6/18)\Gamma(13/18)} 
= \frac{\sin(\pi/9)\sin(4\pi/9)}{2^{-10/9}3^{1/6}\pi}.
$$
$$
\Gamma(E_7,\alpha_7) = \frac{\Gamma(4/18)}{\Gamma(1/18)\Gamma(9/18)\Gamma(12/18)} 
= \frac{\sin(\pi/9)}{2^{-1/9}3^{1/6}\pi}.
$$
Il s'en suit :
$$
\Gamma(E_7) = 2^{1/9}3^{-1/6}\pi^{-1}(2\sin(\pi/9)\sin(2\pi/9),\sin(2\pi/9),\sqrt 3/2, 
2\sin(2\pi/9)\sin(4\pi/9),
$$
$$
2\sin(2\pi/9)\sin(5\pi/18),2\sin(\pi/9)\sin(4\pi/9),\sin(\pi/9)).
$$
En utilisant (M) avec $n = 1/9$, on a: 
$$
\sin(\pi/9)\sin(2\pi/9)\sin(4\pi/9) = \frac{\sqrt 3}{8},
\eqno{(*)}
$$
Il s'en suit que :
$$
\Gamma(E_7) = 2^{1/9}3^{-1/6}\sin(\pi/9)\pi^{-1}m(E_7).
$$

\bigskip

Le cas $E_8$ est plus p\'enible; il est trait\'e dans l'Appendice.

\bigskip

\centerline{\it Syst\`eme de racines $F_4$}

\bigskip

$h = 12$. 

$$
\Gamma(F_4,\alpha_1) = \frac{1}{\Gamma(6/12)}\cdot\frac{\Gamma(2/12)\Gamma(10/12)}{\Gamma(1/12)\Gamma(11/12)}\cdot 
\frac{\Gamma(3/12)}{\Gamma(4/12)\Gamma(5/12)} = \Gamma(E_6,\alpha_2) = 
$$
$$
= 2^{-1/4}3^{1/8}\pi^{-1} (\sqrt 3 + 1)^{-1/2}.
$$
$$
\Gamma(F_4,\alpha_2) = \frac{1}{\Gamma(6/12)\Gamma(2/12)\Gamma(10/12)}\cdot\frac{\Gamma(1/12)\Gamma(4/12)\Gamma(5/12)\Gamma(9/12)}
{\Gamma(3/12)^2\Gamma(7/12)} = \Gamma(E_6,\alpha_4) = 
$$
$$
= 2^{-3/4}3^{1/8}\pi^{-1} (\sqrt 3 + 1)^{1/2}.
$$
$$
\Gamma(F_4,\alpha_3) = \frac{\Gamma(6/12)}{\Gamma(4/12)\Gamma(5/12)\Gamma(9/12)} = 
\Gamma(E_6,\alpha_3) = \Gamma(E_6,\alpha_5) = 2^{-5/4}3^{1/8}\pi^{-1} (\sqrt 3 + 1)^{1/2}.
$$
$$
\Gamma(F_4,\alpha_4) = \frac{\Gamma(3/12)}{\Gamma(1/12)\Gamma(6/12)\Gamma(8/12)} = 
\Gamma(E_6,\alpha_1) = \Gamma(E_6,\alpha_6) = 2^{-5/4}3^{1/8}\pi^{-1} (\sqrt 3 - 1)^{1/2}.
$$
d'o\`u
$$
\Gamma(F_4) = \Gamma(F_4,\alpha_4)(\sqrt 2, \sqrt 3 + 1, (\sqrt 3 + 1)/\sqrt 2, 1)=  
$$
$$
= 2^{-5/4}3^{1/8}\pi^{-1} (\sqrt 3 - 1)^{1/2} m(F_4).
$$

\bigskip

\centerline{\it Syst\`eme de racines $G_2$}

\bigskip

$h = 6$

$$
\Gamma(G_2,\alpha_1) = \frac{\Gamma(1/3)}{\Gamma(1/6)\Gamma(1/2)\Gamma(2/3)} = 
2^{-2/3}\pi^{-1} = \Gamma(D_4,\alpha_1).
$$
En utilisant la formule de duplication  :
$$
\Gamma(1/3) = 2^{-2/3}\pi^{-1/2}\Gamma(1/6)\Gamma(2/3).
$$
Ensuite, 
$$
\Gamma(G_2,\alpha_2) = \frac{\Gamma(1/6)\Gamma(2/3)}{\Gamma(1/3)^3\Gamma(5/6)} = 
$$
en employant la formule des compl\'ements pour $x = 1/6$ et $1/3$, 
$$
= \frac{\sqrt 3}{4\pi^2}\cdot\biggl[\frac{\Gamma(1/6)\Gamma(2/3)}{\Gamma(1/3)}\biggr]^2 = 
\frac{\sqrt 3}{2^{2/3}\pi} = \Gamma(D_4,\alpha_2).
$$
Il s'en suit: 
$$
\Gamma(G_2) = 2^{-2/3}\pi^{-1}(1,\sqrt 3) = 2^{-2/3}\pi^{-1} m(G_2).
$$

\bigskip\bigskip  



\centerline{\bf \S 3. Preuves: cas affine}

\bigskip\bigskip

On utilise toujours les notations de \S 1. 

Supposons d'abord que $R$ soit simpl\'ement lac\'e. 

{\bf 3.1.} {\it Th\'eor\`eme} (V.Fateev, [F], (66)). 

Pour $1\leq i \leq r$,  on a
$$
\gamma(R,\alpha_i) = k(R)^{-1/h}n_i
\eqno{(3.1)} 
$$

L'argument de Fateev est indirect; il vient de l'Ansatz de Bethe. Une preuve directe qui n'utilise que 
la propri\'et\'e (M) de la fonction Gamma est possible, cf. [CA]. 

Ici nous supposons (3.1) connue. 

On a :
$$
\alpha_0 = - \sum_{i=1}^r\ n_i\alpha_i,
$$
donc 
$$
\gamma(R,\alpha_0) = \prod_{i=1}^r\ \gamma(R,\alpha_i)^{-n_i} = k(R)^{\sum_1^r n_i/h} \prod_{i=1}^r\ n_i^{-n_i} = k(R)^{-1/h},
$$
puisque $\sum_1^r n_i = h-1$. Il s'en suit que 
$$
\gamma(R) = k(R)^{-1/h}\delta,
$$
ce qui d\'emontre Th\'eor\`eme 1.2. $\square$. 

R\'eciproquement, la valeur du facteur $k(R)^{-1/h}$ est uniquement d\'efinie si on veut que 
$\gamma(R)$ soit proportionel \`a $\delta$.  

Pour le cas g\'en\'eral (pas forcement simplement lac\'e), posons :
$$
k(R) = \prod_{i=1}^r\ n_i^{\vee n_i}.
$$
La formule de C.Ahn, P.Baseilhac, V.A.Fateev, C.Kim et C.Rim dit :

{\bf 3.2.} {\it Th\'eor\`eme} ([APFKR]). Pour $1\leq i \leq r$,  on a
$$
\gamma(R,\alpha_i) = k(R)^{-1/h}n_i^\vee
\eqno{(3.2)} 
$$

En supposant (3.2) connue, le m\^eme argument comme ci-dessus montre qu'elle implique Th\'eor\`eme 1.3.

\bigskip\bigskip

\newpage

\centerline{\bf \S 4. Autour de sommes de Jacobi}

\bigskip\bigskip

Soient $N$ un entier $\geq 2$, $A_N = N^{-1}\BZ/\BZ$,  
$B_N$ l'anneau de fonctions $f:\ A_N \lra \BQ$. Pour $a\in A_N$, soit $\langle a \rangle$ le r\'epresentant de $a$ dans l'intervale $]0, 1[$. On va utiliser 
la notation suivant: une fonction $f\in B_N$ sera \'ecrit sous une forme
$$
f = \sum_{a\in A_N}\ f(a)[a] = \sum_{i=1}^{N-1}\ f(i/N)[i]
$$
On pose
$$
\tf = \sum_{a\in A_N}\ (f(a) - f(1-a))[a]
$$
Pour $f\in B_N$ on d\'efinit le nombre 
$$
\Gamma(f) = \prod_a\ \Gamma(\langle a \rangle)^{f(a)}
\eqno{(4.1)}
$$
Alors
$$
\Gamma(\tf) = \prod_a\ \gamma(\langle a \rangle)^{f(a)}
\eqno{(4.2)}
$$
Posons
$$
\bbn(f) = \sum_{a\in A_N}\ \langle a \rangle f(a)
$$ 
Le groupe 
$U_N = (\BZ/N\BZ)^*$ agit sur $B_N$ de fa\c{c}on naturelle: $(uf)(a) := f(ua)$.

On d\'efinit un sous-groupe 
$$
C_N = \{f\in B_N|\ \bbn(f)\in\BZ,\ \bbn(f) = \bbn(uf)\ \text{pour tout\ }u\in U_N\}\subset B_N
$$
et pour $k\in\BZ$ 
$$
C_{N,k} = \{f\in C_N|\ \bbn(f) = k\}
$$
  
{\bf 4.1.} {\it Th\'eor\`eme} (Koblitz - Ogus, [KO], Deligne, [D2]). Si $f \in C_{N,k}$  alors $\pi^{-k}\Gamma(f)$ est un nombre alg\'ebrique.

Il serait int\'eressant \`a savoir si la r\'eciproque est vraie, i.e.  
est-il vrai que, \'etant donn\'ee une fonction $f\in B_N$ telle que $\pi^{-k}\Gamma(f)$ 
est alg\`ebrique, alors $f \in C_{N,k}$?

En tout cas, les produits Gamma discut\'es pr\'ecedemment donnent lieu aux telles fonctions. 
En effet, pour $R$ comme ci-dessus, on prend pour $N$ le nombre de Coxeter $h$ de $R$. Pour 
chaque $\alpha > 0$, $(\alpha|\rho)$ est un entier tel que 
$0 < (\alpha|\rho) < h$, et pour tout $1\leq i \leq r$, $(\alpha|\alpha_i)$ est un entier. 
D\'efinissons $f_{R,i}\in B_h$ par 
$$
f_{R,i}(j/h) = - \sum_{\alpha > 0: (\alpha|\rho) = j} (\alpha|\alpha_i)
$$
\'Evidemment, 
$$
\Gamma(f_{R,i}) = \Gamma(R,i)
$$

{\bf 4.2.} {\it Proposition.} Pour tout $R, i$, $f_{R,i} \in C_{N,-1}$ et $\tf_{R,i} \in C_{N,0}$. 

{\it D\'emonstration.} On a calcul\'e les valeurs de ces produits n'en utilisant que les formules 
(C) et (M) du \S 1; ceci implique l'assertion, cf. [KO]. 

Pour les syst\`emes exceptionelles on peut v\'erifier cela directement. $\square$

{\bf 4.3.} {\it Exemple.}   
\'Ecrivons explicitement les fonctions qui correspondent au syst\`eme de racines $E_6$. 

On a $h = 12$. 
Le groupe $U_{12} = \{1, 5, 7, 11\} \cong \BZ/2\BZ \times \BZ/2\BZ$, avec des g\'en\'erateurs $5, 7$. 
$$
f_{E_6,1} = f_{E_6,6} = - [1] + [3] - [6] - [8], 
$$
$$
f_{E_6,2} = - [1] + [2] + [3] - [4] - [5] - [6] + [10] - [11], 
$$
$$
f_{E_6,3} = f_{E_6,5} = - [4] - [5] + [6] - [9], 
$$
$$
f_{E_6,4} = [1] - [2] - 2[3] + [4] + [5] - [6] - [7] + [9] - [10].  
$$

{\bf 4.4.} {\it Corollaire.} Pour tout racine simple $\alpha_i$, 
$$
\sum_{\alpha > 0}\ (\alpha_i|\alpha^\vee)(\rho^\vee|\alpha) = h.
$$

Revenons au cas g\'en\'eral. Suivant Weil et Deligne, on d\'efinit, \`a partir des \'el\'ements 
de $B_N$, des sommes de Jacobi et des caract\`eres de Hecke, [W], [D1], [GK]. 
Rappelons la construction. Soit $K = \BQ(\mu_N)$ o\`u $\mu_N\subset \BC^*$ est le groupe de racines 
$N$-\`emes de $1$. Soient $\fp$ un id\`eal premier de $K$ ne divisant pas $2N$, $k(\fp)$ le corps de r\'esidus, 
$N(\fp) = |k(\fp)| = q = p^f$, $t$ l'isomorphisme 
$$
t:\ \{x\in k^*(\fp) | x^N = 1\} \iso \mu_N
$$
inverse \`a la r\'eduction modulo $\fp$. Fixons un caract\`ere additif
$$
\Psi:\ \BF_p \iso \mu_p.
$$
Pour $a\in A_N$ on d\'efinit la somme de Gauss
$$
g(a,\fp) = - \sum_{x\in k(\fp)^*}\ t(x^{a(q-1)})\Psi(tr(x)) \in K(\mu_p)
$$
Pour $f\in B_N$ on d\'efinit "la somme de Jacobi"
$$
J(f,\fp) = \prod_{a\in A_N}\ g(a,\fp)^{f(a)} 
$$
Quand $\fp$ varie, les nombres $J(f,\fp)$ forment, d'apr\`es Weil, [W], un caract\`ere de Hecke. 

Maintenant supposons que $f\in C_{N,k}$. On d\'efinit le caract\`ere de Hecke de $K$ $\chi_f$ par 
$$
\psi_f(\fp) = N(\fp)^{-k}J(f,\fp); 
$$
il prend ces valeurs dans $K$ et est d'ordre fini. Donc on peut passer du c\^ot\'e automorphe au c\^ot\'e 
galoisien: par la th\'eorie de corps de classes \`a $\psi_f$ correspond un caract\`ere de Galois
$$
\chi(f): \Gal(\bar\BQ/K) \lra K^*
$$
tel que $\psi_f(\fp)$ est \'egal \`a la valeur de $\chi_f$ sur l'\'el\'ement de Frobenius g\'eom\'etrique 
$F_\fp$. 
Voici une forme plus pr\'ecise de 4.1: 

{\bf 4.5.} {\it Th\'eor\`eme} (Deligne, [D2]; Gross-Koblitz, [GK]). Si $f\in C_{N,k}$, alors 
$$
\tilde\Gamma(f) := (2\pi i)^{-k}\Gamma(f)
$$
est alg\'ebrique sur $K$ et 
$$
\sigma\tilde\Gamma(f) = \chi(f)(\sigma)\tilde\Gamma(f)
$$

{\bf 4.6.} Soient $R$ comme ci-dessus, muni d'une base des racines simples 
$B = \{\alpha_1,\ldots,\alpha_r\} \subset R$, $h = h(R)$, $K = \BQ(\mu_h)$. Pour chaque $1 \leq i \leq r$ on obtient les caract\`eres de Galois
$$
\chi(f_{R,i}),\ \chi(\tf_{R,i}):\ \Gal(\bar\BQ/K) \lra K^*
$$
Par contre, pour tous $i, j$, $\Gamma(R,i)/\Gamma(R,j), \gamma(R,i)/\gamma(R,j) \in K$, donc 
ces caract\`eres ne dependent pas de $i$. Nous les notons par $\chi_{R,B}, \tchi_{R,B}$. 

Si $R = A_r, B_r$ ou $C_r$ alors $\chi_{R,B}$ est trivial.  
 
Les bases de $R$ forment un torseur $\CW(R)$ sous le groupe de Weyl $W(R)$, cf. [B], Ch. VI, 1.5, Remarque 4, 
d'o\`u les applications
$$
\chi_{R}:\ \CW(R)\times \Gal(\bar\BQ/K)\lra K^*
$$
et  
$$
\tilde\chi_{R}:\ \CW(R)\times \Gal(\bar\BQ/K)\lra K^*
$$

{\bf 4.7.} 
Les int\'egrales (4.1) apparaissent comme les p\'eriodes de motives de rang $1$ cont\'enus dans la 
cohomology des hypersurfaces de Fermat, cf. [D2]. Il existe une autre source de motives de rang $1$: 
les int\'egrales de Selberg. 

Consid\'erons, avec K.Aomoto, [A], deux types d'int\'egrales de Selberg:

(a) l'int\'egrale de Selberg classique, ou "re\'elle", [S]: 
$$
I_\BR(\alpha,\beta,\rho;n) := \int_{[0,1]^n} \prod_{j=1}^n x_j^{\alpha-1}(1 - x_j)^{\beta-1} 
\prod_{1\leq j < k \leq n} (x_j - x_k)^{2\rho} dx_1\ldots dx_n; 
\eqno{(4.3)}
$$

(b) l'int\'egrale de Dotsenko - Fateev, ou "Selberg complexe", cf. [DF], [A]:
$$
I_\BC(\alpha,\beta,\rho;n) := 
$$
$$
\int_{\BC^n} \prod_{j=1}^n |z_j|^{2(\alpha-1)}|1 - z_j|^{2(\beta-1)} 
\prod_{1\leq j < k \leq n} |z_j - z_k|^{4\rho} \prod_{j=1}^n (i/2)dz_j \bar{dz}_j.  
\eqno{(4.4)}
$$
D'apr\`es Selberg,
$$
I_\BR(\alpha,\beta,\rho;n) = 
\prod_{j=0}^{n-1}\frac{\Gamma(1 + \rho + j\rho)\Gamma(\alpha + j\rho)\Gamma(\beta + j\rho)}
{\Gamma(1 + \rho)\Gamma(\alpha + \beta + (n+j-1)\rho)}
$$
D'un autre c\^ot\'e, le th\'eor\`eme d'Aomoto [A] peut \^etre \'ecrit sous une forme
$$
I_\BC(\alpha,\beta,\rho;n) = \pi^{n}
\prod_{j=0}^{n-1}\frac{\gamma(1 + \rho + j\rho)\gamma(\alpha + j\rho)\gamma(\beta + j\rho)}
{\gamma(1 + \rho)\gamma(\alpha + \beta + (n+j-1)\rho)}
$$
En revenant \`a notre exemple, on voit que le passage d'une matrice de Cartan finie \`a la matrice affine 
est parall\`ele au passage des int\'egrales de Selberg r\'eelles aux int\'egrales complexes.          

\bigskip



\centerline{\bf Appendice. Le cas $E_8$}

\bigskip\bigskip

On va utiliser les formules trigonom\'etriques \'el\'ementaires suivantes. 

$$
s(x) = \frac{\pi}{\sin(\pi x)}
$$
$$
\Gamma(x)^2 = \gamma(x)s(x)
\eqno{(1)}
$$

$$
\sin(3x) = \sin x(4\cos^2x - 1) = \sin x(3 - 4\sin^2x)
\eqno{(2)}
$$

$$
\cos(\pi/5) = \sin(3\pi/10) = \frac{1 + \sqrt 5}{4};\ \frac{1}{\cos(\pi/5)} = \sqrt 5 - 1
\eqno{(3a)}
$$
$$
\cos(2\pi/5) = \sin(\pi/10) = \frac{- 1 + \sqrt 5}{4}
\eqno{(3b)}
$$
$$
\sin^2(\pi/5) = \frac{5 - \sqrt 5}{8}
\eqno{(3c)}
$$
$$
\sin(\pi/5)\sin(2\pi/5) = \frac{\sqrt 5}{4}
\eqno{(3d)}
$$
$$
\sin(2\pi/5) = \frac{\sqrt 5 + 1}{2}\sin(\pi/5)
\eqno{(3e)}
$$

$$
\frac{\sin(3\pi/10)}{\sin(\pi/10)} = 4\cos^2(\pi/5)
\eqno{(4)}
$$

$$
\sin(\pi/15)\sin(4\pi/15) = \frac{1}{2}(\cos(\pi/5) - \cos(\pi/3)) = 
\frac{\sqrt 5 - 1}{8}
\eqno{(5)}
$$

On pose 
$$
\alpha:= \sin(\pi/5).
$$
Alors: 
$$
\sin(2\pi/15) = - \frac{1}{2}\alpha + \frac{1 + \sqrt 5}{4}\cdot \frac{\sqrt 3}{2}
\eqno{(6a)}
$$
$$
\sin(4\pi/15) = \frac{1 + \sqrt 5}{4}\cdot\alpha + \frac{-1 + \sqrt 5}{4}\cdot \frac{\sqrt 3}{2}
\eqno{(6b)}
$$
$$
\sin(8\pi/15) = \frac{1}{2}\alpha + \frac{1 + \sqrt 5}{4}\cdot \frac{\sqrt 3}{2}
\eqno{(6c)}
$$
$$
\sin(\pi/15) =\frac{1 + \sqrt 5}{4}\alpha - \frac{-1 + \sqrt 5}{4}\cdot \frac{\sqrt 3}{2}
\eqno{(6d)}
$$

Ensuite:
$$
\sin(\pi/15)\cdot \sin(4\pi/15) = \frac{-1 + \sqrt 5}{8}
\eqno{(7a)}
$$
$$
\sin(2\pi/15)\cdot \sin(8\pi/15) = \frac{1 + \sqrt 5}{8}
\eqno{(7b)}
$$

$$
\sin(7\pi/30) = \frac{1 - \sqrt 5}{8} + \frac{(1 + \sqrt 5)\sqrt 3}{4}\alpha
\eqno{(8a)}
$$
$$
\sin(11\pi/30) = \frac{1 + \sqrt 5}{8} + \frac{\sqrt 3}{2}\alpha
\eqno{(8b)}
$$
$$
\sin(7\pi/30)\sin(13\pi/30) = \frac{3 + \sqrt 5}{8} = \sin(3\pi/10)^2
\eqno{(8c)}
$$
$$
\sin(4\pi/15)\cdot \sin(8\pi/15) = \sin(3\pi/10)\cdot \sin(11\pi/30)
\eqno{(9)}
$$

\bigskip\bigskip

Les racines positives forment $2$ groupes : 
$$
\alpha(\pm,ij) = \pm \epsilon_i + \epsilon_j,\ 1\leq i < j \leq 8,
$$
$56$ racines ; 
$$
\alpha(\pm,\pm,\ldots) = \frac{1}{2}(\epsilon_8 + \sum_{i=1}^7)\ s_i\epsilon_i,\ s_i = \pm 1,\ \prod s_i = 1,
$$
$64$ racines ; $56 + 64 = 120$ racines positives. 


Le nombre de Coxeter $h = 30$. 

$$
\gamma_F(E_8) := (\gamma(E_8,\alpha_1), \ldots, \gamma(E_8,\alpha_8)) = 
$$
$$
= 2^{-13/15}3^{-2/5}5^{-1/6}(2, 3, 4, 6, 5, 4, 3, 2) = 
$$
$$
= (2^{2/15}3^{-2/5}5^{-1/6}, 2^{-13/15}3^{3/5}5^{-1/6}, 2^{17/15}3^{-2/5}5^{-1/6}, 2^{2/15}3^{3/5}5^{-1/6}, 
$$
$$ 
2^{-13/15}3^{-2/5}5^{5/6}, 2^{17/15}3^{-2/5}5^{-1/6}, 2^{-13/15}3^{3/5}5^{-1/6}, 2^{2/15}3^{-2/5}5^{-1/6}).
$$

$$
m(E_8) = (2\cos(\pi/5), 4\cos(\pi/5)\cos(7\pi/30), 4\cos(\pi/5)\cos(\pi/30), 
$$
$$ 
8\cos^2(\pi/5)\cos(2\pi/15), 8\cos^2(\pi/5)\cos(7\pi/30), 4\cos(\pi/5)\cos(2\pi/15), 
2\cos(\pi/30), 1) = 
$$
$$
= (1,62; 2,40; 3,22; 4,78; 3,89; 2,96; 1,99; 1).
$$
--- les masses des particules dans le mod\`ele d'Ising critique avec le champs magn\'etique (Zamolodchikov). 

$$
\gamma(E_8,\alpha_1) = \frac{\gamma(3/30)\gamma(5/30)\gamma(16/30)}
{\gamma(1/30)\gamma(8/30)\gamma(10/30)\gamma(12/30)\gamma(23/30)} = 
$$
$$
= 2^{2/15}3^{-2/5}5^{-1/6}.
$$

$$
\Gamma(E_8,\alpha_1) = \frac{\Gamma(3/30)\Gamma(5/30)\Gamma(16/30)}
{\Gamma(1/30)\Gamma(8/30)\Gamma(10/30)\Gamma(12/30)\Gamma(23/30)} =   
$$
$$
2^{16/15}3^{1/20}5^{-1/12}\pi^{-1}\sin(\pi/15)\cdot\biggl(
\frac{\sin(2\pi/5)\sin(2\pi/15)\sin(4\pi/15)}{\sin(\pi/10)}\biggr)^{1/2}.
$$

$$
\gamma(E_8,\alpha_2) = \frac{\gamma(2/30)\gamma(3/30)\gamma(10/30)\gamma(12/30)\gamma(21/30)}
{\gamma(1/30)\gamma(6/30)\gamma(7/30)\gamma(8/30)\gamma(15/30)\gamma(17/30)\gamma(24/30)} = 
$$
$$
= 2^{-13/15}3^{3/5}5^{-1/6}.
$$

$$
\Gamma(E_8,\alpha_2) = \frac{\Gamma(2/30)\Gamma(3/30)\Gamma(10/30)\Gamma(12/30)\Gamma(21/30)}
{\Gamma(1/30)\Gamma(6/30)\Gamma(7/30)\Gamma(8/30)\Gamma(15/30)\Gamma(17/30)\Gamma(24/30)}  
$$
$$
2^{-103/30}3^{1/20}5^{-1/12}\pi^{-1}\frac{\sin(\pi/5)}{\sin(\pi/15)}\cdot 
[\sin(4\pi/15)\sin(8\pi/15)\sin(\pi/10)\sin(3\pi/10)\sin(2\pi/5)]^{-1/2}.
$$
En utilisant (7a) et (7b), on obtient : 
$$
\Gamma(E_8,\alpha_2)/\Gamma(E_8,\alpha_8) = \frac{1}{2\sin(\pi/15)}.
$$
D'un autre c\^ot\'e, 
$$
m_2 = m_2/m_8 = 4\cos(\pi/5)\sin(4\pi/15) = (1 + \sqrt 5)\sin(4\pi/15),
$$
d'o\`u :
$$
\Gamma(E_8,\alpha_2)/\Gamma(E_8,\alpha_8) =  m_2/m_8.
$$

$$
\gamma(E_8,\alpha_3) 
= \frac{\gamma(8/30)\gamma(15/30)\gamma(22/30)}{\gamma(7/30)\gamma(11/30)\gamma(13/30)
\gamma(20/30)\gamma(24/30)} =   
$$
$$
= 2^{17/15}3^{-2/5}5^{-1/6}.
$$

$$
\Gamma(E_8,\alpha_3) 
= \frac{\Gamma(8/30)\Gamma(15/30)\Gamma(22/30)}{\Gamma(7/30)\Gamma(11/30)\Gamma(13/30)
\Gamma(20/30)\Gamma(24/30)} =   
$$
$$
= 2^{1/15}3^{1/20}5^{-1/12}\pi^{-1}\sin(4\pi/15)^{-1}\bigl(\sin(\pi/5)
\sin(7\pi/30)\sin(11\pi/30)\sin(13\pi/30)\bigr)^{1/2} = 
$$
$$
= 2^{17/30}3^{1/20}5^{-1/12}\pi^{-1}\sin(8\pi/15)\cdot\biggl(
\frac{\sin(\pi/5)\sin(2\pi/15)}{\sin(4\pi/15)}\biggr)^{1/2}.
$$

De l\`a, on obtient : 
$$
\Gamma(E_8,\alpha_3)/\Gamma(E_8,\alpha_8) = 2^{-1/2}(1 + \sqrt 5)\sin(3\pi/10)\cdot \biggl(\frac{\sin(11\pi/30)}{\sin(2\pi/15)\sin(4\pi/15)}\biggr)^{1/2}.
$$
D'autre part : 
$$
m_3 = 4\sin(3\pi/10)\sin(8\pi/15),
$$
d'o\`u : 
$$
\Gamma(E_8,\alpha_3)/\Gamma(E_8,\alpha_8) = m_3,
$$
en utilisant (9). 

$$
\gamma(E_8,\alpha_4) 
= \frac{\gamma(1/30)\gamma(4/30)\gamma(11/30)\gamma(20/30)\gamma(24/30)}{\gamma(2/30)
\gamma(3/30)\gamma(8/30)\gamma(12/30)\gamma(18/30)\gamma(22/30)\gamma(25/30)} =   
$$
$$
= 2^{2/15}3^{3/5}5^{-1/6}.
$$

$$
\Gamma(E_8,\alpha_4) 
= \frac{\Gamma(1/30)\Gamma(4/30)\Gamma(11/30)\Gamma(20/30)\Gamma(24/30)}{\Gamma(2/30)
\Gamma(3/30)\Gamma(8/30)\Gamma(12/30)\Gamma(18/30)\Gamma(22/30)\Gamma(25/30)} =     
$$
$$
= 2^{-14/15}3^{1/20}5^{-1/12}\pi^{-1}\sin(2\pi/5)\cdot\biggl(
\frac{\sin(\pi/10)}{\sin(\pi/5)\sin(\pi/15)\sin(2\pi/15)}\biggr)^{1/2}.
$$
On a : 
$$
m_4 = 8\cdot\frac{\sin(3\pi/10)}{\sin(4\pi/15)\sin(8\pi/15)}
$$
(on utilise (9)). Il s'en suit, en employant (7) et (3), que : 
$$
\Gamma(E_8,\alpha_4)/\Gamma(E_8,\alpha_8) = m_4.
$$

$$
\gamma(E_8,\alpha_5) = 
$$
$$ 
\frac{\gamma(2/30)\gamma(6/30)\gamma(7/30)\gamma(8/30)\gamma(12/30)\gamma(13/30)\gamma(18/30)
\gamma(25/30)}{\gamma(4/30)^2\gamma(5/30)\gamma(9/30)\gamma(10/30)\gamma(11/30)\gamma(15/30)\gamma(16/30)\gamma(21/30)\gamma(26/30)} =  
$$
$$
= 2^{-13/15}3^{-2/5}5^{5/6}.
$$

$$
\Gamma(E_8,\alpha_5) = 
$$
$$ 
\frac{\Gamma(2/30)\Gamma(6/30)\gamma(7/30)\Gamma(8/30)\Gamma(12/30)\Gamma(13/30)\Gamma(18/30)
\Gamma(25/30)}{\Gamma(4/30)^2\Gamma(5/30)\Gamma(9/30)\Gamma(10/30)\Gamma(11/30)\Gamma(15/30)\Gamma(16/30)\Gamma(21/30)\Gamma(26/30)} =    
$$
$$
2^{-13/30}3^{1/20}5^{5/12}\pi^{-1}\frac{\sin(3\pi/10)}{\sin(2\pi/5)}\cdot 
\biggl(\frac{\sin(2\pi/15)\sin(4\pi/15)}{\sin(\pi/5)}\biggr)^{1/2}.
$$
En utilisant (3d), (4) et (5), on v\'erifie sans peine que : 
$$
\Gamma(E_8,\alpha_5)/\Gamma(E_8,\alpha_8) = 8\cos^2(\pi/5)\cos(2\pi/15) = 
m_5/m_8.
$$

$$
\gamma(E_8,\alpha_6) 
= \frac{\gamma(9/30)\gamma(15/30)\gamma(26/30)}{\gamma(6/30)\gamma(13/30)\gamma(14/30)
\gamma(20/30)\gamma(27/30)} =   
$$
$$
= 2^{17/15}3^{-2/5}5^{-1/6}.
$$

$$
\Gamma(E_8,\alpha_6) 
= \frac{\Gamma(9/30)\Gamma(15/30)\Gamma(26/30)}{\Gamma(6/30)\Gamma(13/30)\Gamma(14/30)
\Gamma(20/30)\Gamma(27/30)} =     
$$
$$
= 2^{47/30}3^{1/20}5^{-1/12}\pi^{-1}\sin(4\pi/15)\cdot\biggl(
\sin(\pi/5)\sin(\pi/15)\sin(8\pi/15)\biggr)^{1/2}.
$$
On a : 
$$
m_6 = 4\sin(4\pi/15)\sin(8\pi/15)
$$
(en employant (9)), d'o\`u, en utilisant (3c) et (7), 
$$
\Gamma(E_8,\alpha_6)/\Gamma(E_8,\alpha_8) = m_6.
$$

$$
\gamma(E_8,\alpha_7) 
= \frac{\gamma(4/30)\gamma(10/30)\gamma(14/30)\gamma(27/30)}{\gamma(2/30)\gamma(9/30)\gamma(12/30)\gamma(15/30)\gamma(19/30)\gamma(28/30)} =  
$$
$$
= 2^{-13/15}3^{3/5}5^{-1/6}.
$$

$$
\Gamma(E_8,\alpha_7) 
= \frac{\Gamma(4/30)\Gamma(10/30)\Gamma(14/30)\Gamma(27/30)}{\Gamma(2/30)\Gamma(9/30)\Gamma(12/30)\Gamma(15/30)\Gamma(19/30)\Gamma(28/30)} =    
$$
$$
= 2^{-13/30}3^{1/20}5^{-1/12}\pi^{-1}\cdot\biggl(
\frac{\sin(2\pi/5)\sin(\pi/15)}{\sin(2\pi/15)}\biggr)^{1/2}.
$$
En utlisant (7a), 
$$
\Gamma(E_8,\alpha_7)/\Gamma(E_8,\alpha_8) = \frac{1 + \sqrt 5}{4\sin(2\pi/15)}.
$$
D'un autre c\^ot\'e, 
$$
m_7 = 2\sin(8\pi/15),
$$
d'o\`u
$$
\Gamma(E_8,\alpha_7)/\Gamma(E_8,\alpha_8) = m_7, 
$$
vu (7b).

$$
\gamma(E_8,\alpha_8) 
= \frac{\gamma(5/30)\gamma(9/30)\gamma(28/30)}{\gamma(1/30)\gamma(10/30)\gamma(14/30)\gamma(18/30)
\gamma(29/30)} =   
$$
$$
= 2^{2/15}3^{-2/5}5^{-1/6}.
$$

$$
\Gamma(E_8,\alpha_8) 
= \frac{\Gamma(5/30)\Gamma(9/30)\Gamma(28/30)}{\Gamma(1/30)\Gamma(10/30)\Gamma(14/30)\Gamma(18/30)
\Gamma(29/30)} =    
$$
$$
2^{16/15}3^{1/20}5^{-1/12}\pi^{-1}\sin(\pi/15)\cdot\biggl(
\frac{\sin(2\pi/5)\sin(2\pi/15)\sin(4\pi/15)}{\sin(3\pi/10)}\biggr)^{1/2}.
$$
En employant (4), on voit que 
$$
\Gamma(E_8,\alpha_1)/\Gamma(E_8,\alpha_8) = 2\cos(\pi/5) = m_1/m_8.
$$

On a donc v\'erifi\'e  que : 
$$
\Gamma(E_8) = \Gamma(E_8,\alpha_8)m(E_8).
$$

\bigskip\bigskip

\bigskip

\centerline{\bf Bibliographie}

\bigskip\bigskip

[ABFKR] C.Ahn, P.Baseilhac, V.A.Fateev, C.Kim, C.Rim, Reflection amplitudes in non-simply laced 
Toda theories and thermodynamic Bethe Ansatz, {\it Phys. Let.} {\bf B481} (2000), 114 - 124.

[A] K.Aomoto, On the complex Selberg integral, {\it Quart. J. Math. Oxford} (2), {\bf 38} (1987), 385 - 399. 
 
[B] N.Bourbaki, Groupes et alg\`ebres de Lie, Ch. IV - VI. 

[BCDS] H.W.Braden, E.Corrigan, P.E.Dorey, R.Sasaki, Affine Toda field theory and exact $S$-matrices, 
{\it Nucl. Phys.} {\bf B338} (1990), 689 - 746. 

[CA] V.Cohen-Aptel, Formule de Fateev, arXiv:1012.5203.

[D1] P.Deligne, Valeurs de fonctions $L$ et p\'eriodes des int\'egrales, {\it Proc. Symp. Pure Math.} 
{\bf 33} (1979), part 2, 313 - 346. 

[D2] P.Deligne, Hodge cycles on Abelian varieties, dans: P.Deligne, J.S.Milne, A.Ogus, K.Shih, 
Hodge cycles, Motives and Shimura varieties, {\it Lect. Notes Math.} {\bf 900} (1982), 9 - 100. 

[DF] Vl.S.Dotsenko, V.A.Fateev, Four point correlation functions and the operator algebra in 2D conformal 
invariant theories with central charge $c\leq 1$, {\it Nucl. Phys.} {\bf B251} (1985), 691 - 734.  

[F1] V.A.Fateev, Normalization factors, reflection amplitudes and integrable systems, hep-th/0103014. 

[F2] V.A.Fateev, The exact relations between the coupling constants and the masses of particles 
for the integrable perturbed Conformal Field Theories, {\it Phys. Let.} {\bf B324} (1994), 45 - 51. 

[FZ] V.A.Fateev, A.B.Zamolodchikov, Conformal Field Theory and purely elastic $S$-matrices, {\it Int. J. Mod. Phys.} A, {\bf 5} 
(1990), 1025 - 1048. 

[Fr] M.D.Freeman, On the mass spectrum of affine Toda field theory, {\it Phys. Let.} {\bf B261} (1991), 
57 - 61. 

[FLO] A.Fring, H.C.Liao, D.I.Olive, The mass spectrum and coupling in affine Toda theories, 
{\it Phys. Let.} {\bf B266} (1991), 82 - 86. 

[G] F.R.Gantmacher, Th\'eorie des matrices. 

[GHJ] F.M.Goodman, P.de la Harpe, V.F.R.Jones, Coxeter graphs and towers of algebras, MSRI Publ. {\bf 14}, 
Springer, 1989. 

[GK] B.Gross, N.Koblitz, Gauss sums and $p$-adic $\Gamma$-function, {\it Ann. Math.}, {\bf 109} 
(1979), 569 - 581. 

[K] V.G.Kac, Infinite dimensional Lie algebras. 

[KM] T.R.Klassen, E.Melzer, Purely elastic scattering theories and their ultraviolet limits, 
{\it Nucl. Phys.} {\bf B338} (1990), 485 - 528. 

[KO] N.Koblitz, A.Ogus, Algebraicity of some products of values of the $\Gamma$-function, Annexe \`a [D]. 


[P] V.Pasquier, Two dimensional critical systems labelled by Dynkin diagrams, {\it Nucl. Phys.} {\bf B285} 
(1987), 162 - 172. 

[S] A.Selberg, Bemerkninger om et multiplet integral, {\it Norsk. Mat. Tidscr.} {\bf 26} (1944), 71 - 78. 

[W] A.Weil, Jacobi sums as Gr\"ossencharctere, {\it Trans. AMS} {\bf 73} (1952), 487 - 495. 

[Z] A.B.Zamolodchikov, Integrals of motion and $S$-matrix of the (scaled) $T = T_c$ Ising model with magnetic field, {\it Int. J. Mod. Phys.} A, {\bf 4} (1989), 4235 - 4248.

\bigskip\bigskip

Institut de Math\'ematiques de Toulouse, 
Univ\'ersit\'e Paul Sabatier, 118 route de Narbonne, 31062 Toulouse

\end{document}